\documentclass[5p,times,twocolumn,number,sort&compress]{elsarticle}

\usepackage{lineno}
\modulolinenumbers[5]

\journal{Journal of \LaTeX\ Templates}

\usepackage{graphicx}  % needed for figures
\usepackage{amsmath}   % for math
\usepackage{amssymb}   % for math
\usepackage{slashed}
\usepackage[caption=false]{subfig}
\usepackage{tabu}
\usepackage[colorlinks=true,
            urlcolor=blue,
            anchorcolor=blue,
            citecolor=blue,
            filecolor=blue,
            linkcolor=blue,
            menucolor=blue,
            linktocpage=true,
            pdfproducer=medialab,
            pdfa=true,
            bookmarks=false]{hyperref}
\usepackage{paralist}
\usepackage{multirow}
\usepackage{etoolbox}
\apptocmd{\thebibliography}{\raggedright}{}{}
\usepackage[capitalize]{cleveref}

\usepackage{capt-of}

\def\e{\ensuremath{\epsilon}}
\def\d{\ensuremath{\partial}}
\def\gonehhh{\ensuremath{g^{(1)}_{_{hhh}}}}
\def\gtwohhh{\ensuremath{g^{(2)}_{_{hhh}}}}
\def\gonehww{\ensuremath{g^{(1)}_{_{hWW}}}}
\def\gtwohww{\ensuremath{g^{(2)}_{_{hWW}}}}
\def\gonehhww{\ensuremath{g^{(1)}_{_{hhWW}}}}
\def\gtwohhww{\ensuremath{g^{(2)}_{_{hhWW}}}}
\def\gthww{\ensuremath{\tilde g_{_{hWW}}}}
\def\gthhww{\ensuremath{\tilde g_{_{hhWW}}}}
\def\abinv{\ensuremath{\rm ab^{-1}}}

\makeatletter
 \def\@textbottom{\vskip \z@ \@plus 4pt}
 \let\@texttop\relax
\makeatother

\begin{document}

\begin{frontmatter}

\title{Probing anomalous couplings using di-Higgs production in electron-proton collisions}

% \address{Radarweg 29, Amsterdam}
% \fntext[myfootnote]{Since 1880.}

\author[NITheP]{Mukesh Kumar}
\ead{mukesh.kumar@cern.ch}

\author[Wits]{Xifeng Ruan}
\ead{xifeng.ruan@cern.ch}

\author[CU]{Rashidul Islam}
\ead{rashidul.islam@cern.ch}

\author[NITheP]{Alan S. Cornell}
\ead{alan.cornell@wits.ac.za}

\author[Liv]{Max Klein}
\ead{Max.Klein@liverpool.ac.uk}

\author[Liv]{Uta Klein}
\ead{Uta.Klein@liverpool.ac.uk}

\author[Wits]{Bruce Mellado}
\ead{bruce.mellado.garcia@cern.ch}

\address[NITheP]{National Institute for Theoretical Physics,\\
             School of Physics and Mandelstam Institute for Theoretical Physics,\\
             University of the Witwatersrand, Johannesburg, Wits 2050, South Africa.}
\address[Wits]{University of the Witwatersrand, School of Physics,
             Private Bag 3, Wits 2050, South Africa.}
\address[CU]{Department of Physics, University of Calcutta,\\
             92, Acharya Prafulla Chandra Road, Kolkata 700009, India.}
\address[Liv]{Oliver Lodge Laboratory, University of Liverpool,
             Liverpool, United Kingdom.}

% \author[mysecondaryaddress]{Global Customer Service\corref{mycorrespondingauthor}}
% \cortext[mycorrespondingauthor]{Corresponding author}
% \address[mymainaddress]{1600 John F Kennedy Boulevard, Philadelphia}
% \address[mysecondaryaddress]{360 Park Avenue South, New York}

\begin{abstract}
 A proposed high energy Future Circular Hadron-Electron Collider would provide
 sufficient energy in a clean environment to probe di-Higgs production. Using
 this channel we show that the azimuthal angle correlation between the missing
 transverse energy and the forward jet is a very good probe for the non-standard
 $hhh$ and $hhWW$ couplings. We give the exclusion limits on these couplings as
 a function of integrated luminosity at a $95$\% C.L. using the fiducial cross
 sections. With appropriate error fitting methodology we find that the Higgs boson
 self coupling could be measured to be $\gonehhh = 1.00^{+0.24(0.14)}_{-0.17(0.12)}$
 of its expected Standard Model value at $\sqrt s = 3.5(5.0)$~TeV for an ultimate
 $10~\abinv$ of integrated luminosity.
\end{abstract}

\begin{keyword}
Higgs boson, Extensions of Higgs sector, Particle production
\end{keyword}

\end{frontmatter}

% \linenumbers

%%%%%%%%%%%%%%%%%%%%%%%%%%%%%%%%%%%%%%%%%%%%%%%%%%%%%%%%%%%%%%%%%%%%%%%%%%%%%%%
\section{Introduction}
\label{sec:intro}
The $125$~GeV particle discovered by the ATLAS and CMS experiments~\citep{
Aad:2012tfa,Chatrchyan:2012xdj,Aad:2014aba,Khachatryan:2014jba,Aad:2015zhl} has
been established as a spin-0 Higgs boson rather than a spin-2 particle~\citep{
Chatrchyan:2012jja,Aad:2013xqa}. The measurements of its couplings to fermions
and gauge bosons are being updated constantly and the results confirm consistency
with the expected Standard Model (SM) values~\citep{Khachatryan:2014jba,
Khachatryan:2014kca,Aad:2015mxa,Aad:2015gba}. However, to establish that a scalar
doublet $\Phi$ does indeed break the electroweak (EW) symmetry spontaneously when it
acquires a nonzero vacuum expectation value, $v$, requires a direct measurement of 
the Higgs boson self coupling, $\lambda$. The minimal SM, merely on the basis
of the economy of fields and interactions, assumes the existence of only one
physical scalar, $h$, with $J^{PC} = 0^{++}$. Although Ref.~\citep{Khachatryan:2014kca}
has ruled out a pure pseudoscalar hypothesis at a $99.98$\% confidence limit (C.L.),
the new particle can still have a small CP-odd admixture to its couplings.

Theoretically, the Higgs boson self coupling appears when, as a result of
electroweak symmetry breaking in the SM, the scalar potential $V(\Phi)$ gives
rise to the Higgs boson self interactions as follows:
\begin{gather}
 V(\Phi)
 = \mu^2 \Phi^\dag \Phi + \lambda (\Phi^\dag \Phi)^2
 \to \frac{1}{2}m^2_h h^2 + \lambda v h^3 + \frac{\lambda}{4} h^4,
 \label{Vphi}
\end{gather}
where $\lambda \!=\! \lambda_{\rm SM} \!=\! m^2_h/(2v^2) \!\approx\! 0.13$ and $\Phi$ is an
$SU(2)_L$ scalar doublet. For a direct and independent measurement of $\lambda$
we need to access double Higgs boson production experimentally. However, this
path is extremely challenging and requires a very high integrated luminosity to
collect a substantial di-Higgs event rate, and an excellent detector with powerful
background rejection capabilities. On the theoretical side we need to also take into
account all vertices involved in the process that are sensitive to the
presence of new physics beyond the SM (BSM).

There are various proposals to build new, powerful high energy $e^+e^-\!\!$, $e^-p$
and $pp$ colliders in the future. We have based our study on a {\em Future Circular
Hadron-Electron Collider} ({FCC-he}) which employs the $50$~TeV proton beam of a
proposed $100$~km circular $pp$ collider ({FCC-pp}), and electrons from an Energy
Recovery Linac (ERL) being developed for the {\em Large Hadron Electron Collider}
({LHeC})~\citep{AbelleiraFernandez:2012cc,Bruening:2013bga}. The design of the
ERL is such that the $e^-p$ and $pp$ colliders operate simultaneously, thus
optimising the cost and the physics synergies between $e^-p$ and $pp$ processes.
Such facilities would be potent Higgs research centres, see e.g. Ref.~\citep{AbelleiraFernandez:2012ty}. 
The LHeC and the FCC-he configuration are advantageous with respect to the Large Hadron
collider (LHC) (or FCC-pp in general) in terms of
\begin{inparaenum}[(1)]
 \item initial states are asymmetric and hence backward and forward scattering can
 be disentangled,
 \item it provides a clean environment with suppressed backgrounds from strong
 interaction processes and free from issues like pile-ups, multiple interactions etc.,
 \item such machines are known for high precision measurements of the dynamical 
 properties of the proton allowing simultaneous test of EW and QCD effects.
\end{inparaenum}
A detailed report on the physics and detector design concepts can be found in the 
Ref.~\citep{AbelleiraFernandez:2012cc}.

The choice of an ERL energy of $E_e = 60$ to $120$~GeV, with an available proton
energy $E_p = 50\,(7)$~TeV, would provide a centre of mass (c.m.s.) energy of
$\sqrt{s} \approx 3.5(1.3)$ to $5.0(1.8)$~TeV at the FCC-he (LHeC) using
the FCC-pp (LHC) protons. The FCC-he would have sufficient c.m.s. energy to probe
the Higgs boson self coupling via double Higgs boson production. The inclusive
Higgs production cross section at the FCC-he is expected to be about five times
larger than at the proposed $100$~km circular $e^+e^-$ collider (FCC-ee).

This article is organised as follows: We discuss the process to produce the di-Higgs
events in an $e^-p$ collider and the most general Lagrangian with all relevant new
physics couplings in \cref{sec:formalism}. In \cref{sec:analysis} all the simulation
tools and the kinematic cuts that are required to study the sensitivity of the
involved couplings are given. Here we also discuss the details of the analyses
that has gone into the study. In \cref{eftval} there is a discussion on the
validity of the effective theory considered here. And finally we conclude and
draw inferences from the analysis in \cref{sec:conc}.

%%%%%%%%%%%%%%%%%%%%%%%%%%%%%%%%%%%%%%%%%%%%%%%%%%%%%%%%%%%%%%%%%%%%%%%%%%%%%%%
\section{Formalism}
\label{sec:formalism}
\begin{figure}[!htbp]
  \centering
  \subfloat[]{\includegraphics[trim=0 0 0 10,clip,width=0.15\textwidth]{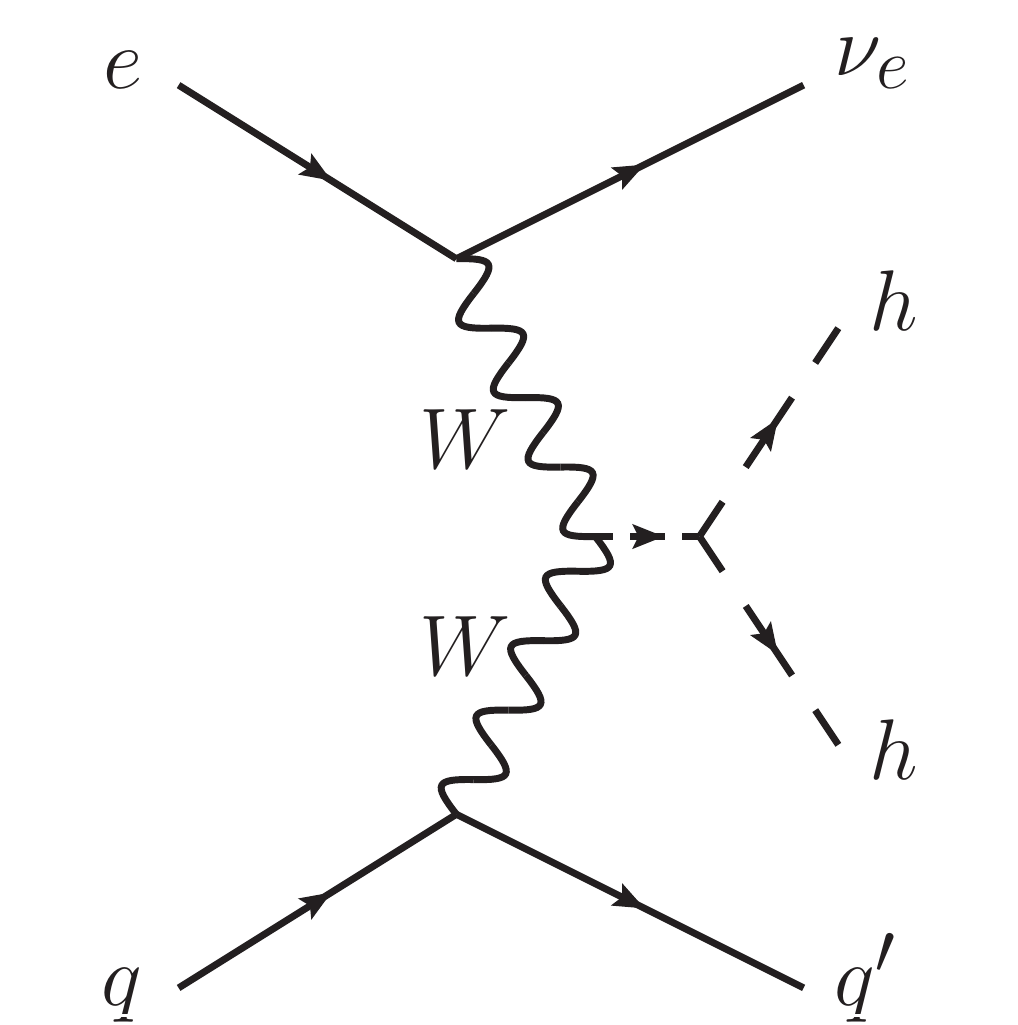}\label{fig:figWc}}
  \subfloat[]{\includegraphics[trim=0 0 0 10,clip,width=0.15\textwidth]{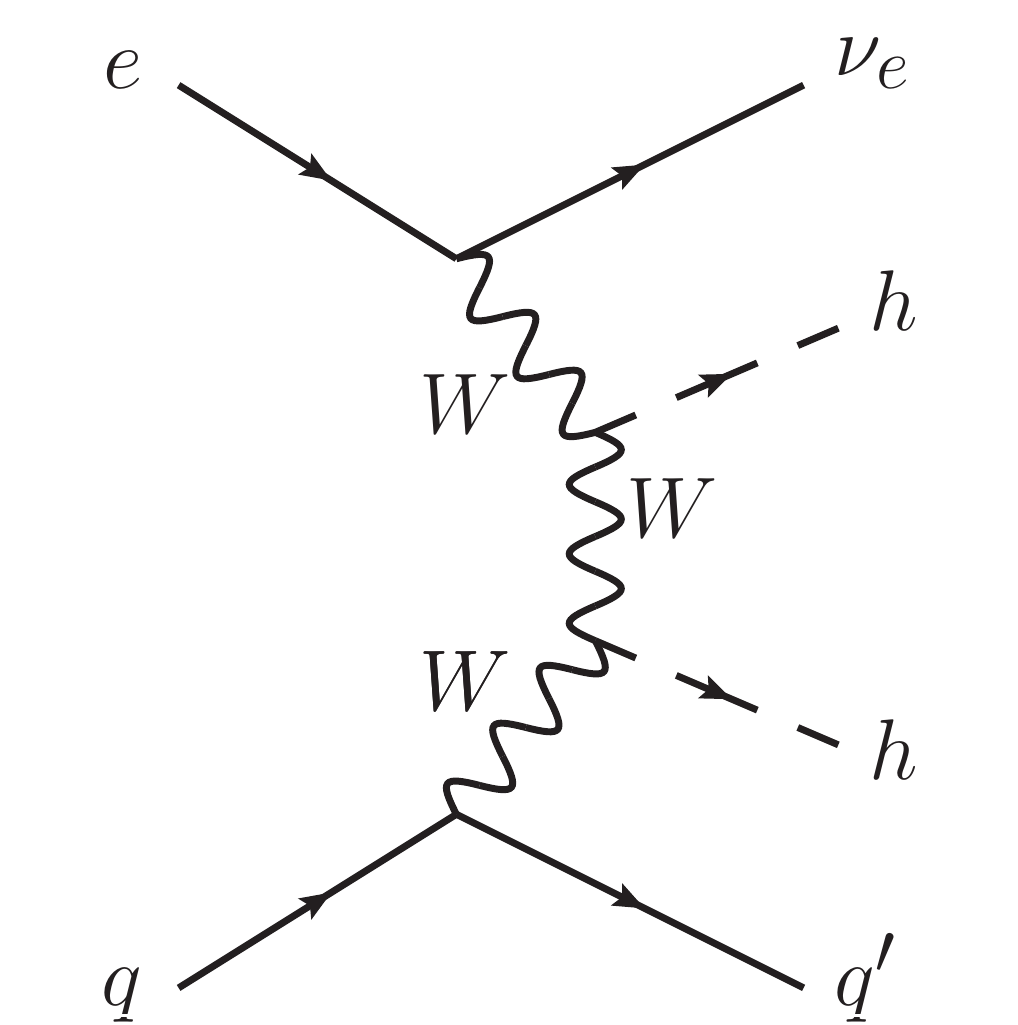}\label{fig:figWb}}
  \subfloat[]{\includegraphics[trim=0 0 0 10,clip,width=0.15\textwidth]{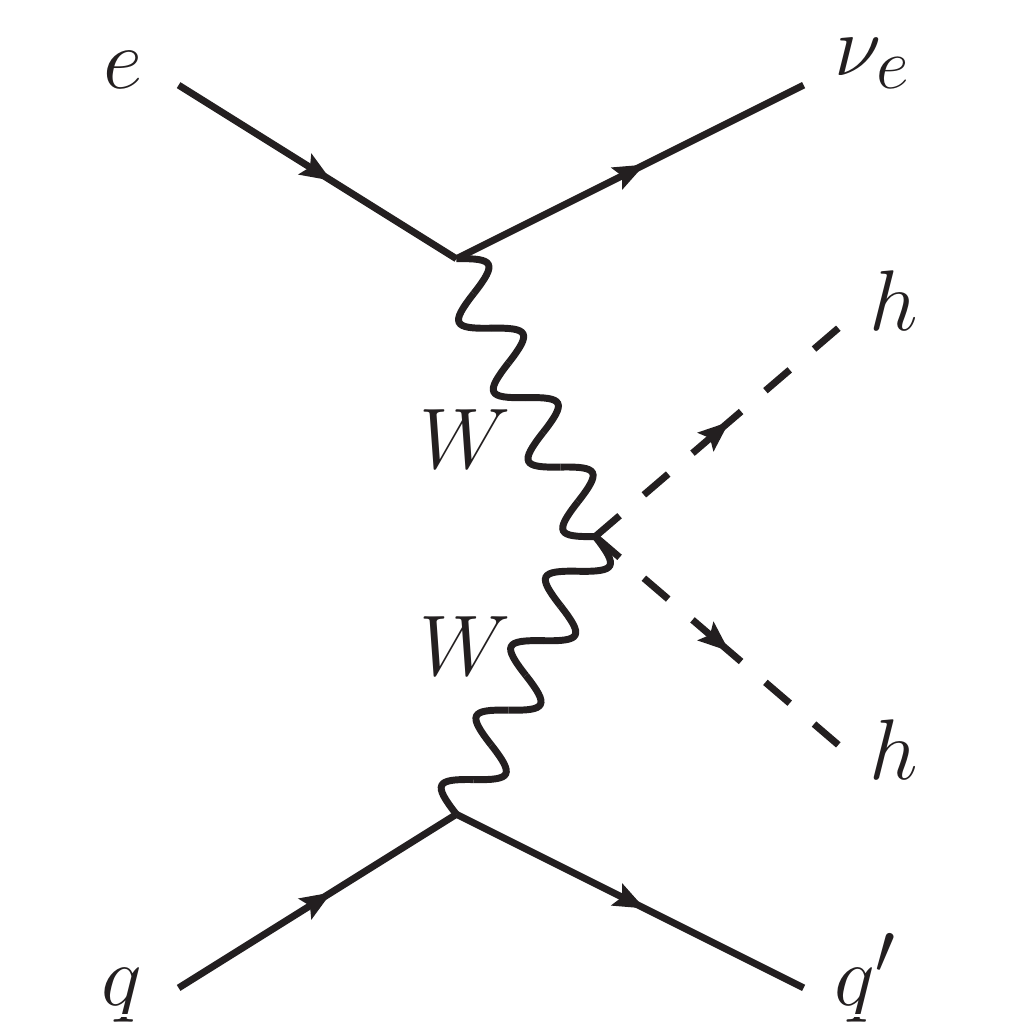}\label{fig:figWa}}
  \caption{\small Leading order diagrams contributing to the process $p\,e^- \to h h j \nu_e$ with
   $q \equiv u,c,\bar d,\bar s$ and $q^\prime \equiv d,s,\bar u, \bar c$ respectively.}
\label{fig:figW}
\end{figure}

In an $e^-p$ collider environment a double Higgs event can be produced through:
\begin{inparaenum}[(1)]
 \item the charged current process, $p\,e^- \to hhj\nu_e$, and
 \item the neutral current process, $p\,e^- \to hhj\,e^-$,
\end{inparaenum}
if there are no new physics processes involved. The SM background will cloud each
of the processes greatly, and it will be a formidable task to separate signal from
backgrounds. Here we study the charged current process because the signal
strength of this is superior to the neutral current process. Hence we show in
\cref{fig:figW} the Higgs boson pair production, at leading order, due to the
resonant and non-resonant contributions in charged current deep inelastic scattering
(CC DIS) at an $e^-p$ collider. As seen in \cref{fig:figW}, the
di-Higgs production involves $hhh, hWW$ and $hhWW$ couplings. Note that the $hWW$
coupling will be extensively probed at the LHC, where its value conforms to the
value predicted by the SM~\citep{Khachatryan:2014jba,Khachatryan:2014kca,Aad:2015gba}.
Through vector boson fusion Higgs production mode at the LHC, a BSM analyses to
determine the CP and spin quantum numbers of the Higgs boson has been studied in
Refs.~\citep{Plehn:2001nj, Hagiwara:2009wt, Djouadi:2013yb, Englert:2012xt}. The
authors of Ref.~\citep{Biswal:2012mp} have shown the sensitivity of new physics
contributions in $hWW$ couplings at $e^-p$ colliders through a study of the
azimuthal angle correlation for single Higgs boson production in $p\,e^- \to
hj\nu_e$ with an excellent signal-to-background ratio based on the $h\to b\bar{b}$
decay channel. Since we do not have any direct measurement of the Higgs boson
self coupling ($hhh$) and quartic ($hhWW$) coupling, there can be
several possible sources of new physics in the scalar sector. This article studies
for the proposed FCC-he sensitivity of the Higgs boson self coupling around
its SM value including BSM contributions by considering all possible Lorentz
structures. In order to make it a complete study we also retain the possibilities
for $hWW$ couplings that appear in the di-Higgs production modes.

Following Refs.~\citep{Biswal:2012mp,Alloul:2013naa} the most general Lagrangian
which can account for all bosonic couplings relevant for the phenomenology of the
Higgs boson sector at the FCC-he are the three-point and four-point interactions
involving at least one Higgs boson field. It can be written as:
\begin{align}
 {\cal L}^{(3)}_{_{hhh}}
 =
   \dfrac{m^2_h}{2v} & (1 - \gonehhh) h^3
 + \dfrac{1}{2 v} \gtwohhh h \d_\mu h \d^\mu h,
 \label{lagh}
 \\
 {\cal L}^{(3)}_{_{hWW}}
 =
 - g \bigg[ & \dfrac{\gonehww}{2m_W} W^{\mu\nu} W^\dag_{\mu\nu} h
 + \dfrac{\gtwohww}{m_W} ( W^\nu \d^\mu W^\dag_{\mu\nu} h + {\rm h.c} )
 \notag\\
 &\quad+  \dfrac{\gthww}{2m_W} W^{\mu\nu} \widetilde W^\dag_{\mu\nu} h \bigg],
 \label{lag3}
 \\
 {\cal L}^{(4)}_{_{hhWW}}
 =
 - g^2 \bigg[ & \dfrac{\gonehhww}{4m^2_W} W^{\mu\nu} W^\dag_{\mu\nu} h^2
 + \dfrac{\gtwohhww}{2m^2_W} ( W^\nu \d^\mu W^\dag_{\mu\nu} h^2 + {\rm h.c} )
 \notag\\
 &\quad+  \dfrac{\gthhww}{4m^2_W} W^{\mu\nu} \widetilde W^\dag_{\mu\nu} h^2 \bigg].
 \label{lag4}
\end{align}
Here $g^{(i)}_{(\cdots)}, i = 1,2$, and $\tilde g_{(\cdots)}$ are real coefficients
corresponding to the CP-even and CP-odd couplings respectively (of the $hhh$, $hWW$
and $hhWW$ anomalous vertices), $W_{\mu\nu} = \d_\mu W_\nu - \d_\nu W_\mu$ and
$\widetilde W_{\mu\nu} = \frac{1}{2} \e_{\mu\nu\rho\sigma} W^{\rho\sigma}$. In
\cref{lagh} $\gonehhh$ is parametrised with a multiplicative constant with respect
to $\lambda_{\rm SM}$ as in \cref{Vphi}. Thus the Higgs self coupling $\lambda$
appears as $\gonehhh \lambda_{\rm SM}$ in the expression for $V(\Phi)$. Clearly,
in the SM $\gonehhh = 1$ and all other anomalous couplings vanish in \cref{lagh,lag3,lag4}.
The Lorentz structures of \cref{lagh,lag3,lag4} can be derived from the $SU(2)_L
\otimes U(1)_Y$ gauge invariant dimension-6 operators given in Ref.~\citep{Alloul:2013naa}.

The complete Lagrangian we work with is as follows:
\begin{align}
 {\cal L}
 =&
 {\cal L}_{\rm SM} + {\cal L}^{(3)}_{_{hhh}}
    + {\cal L}^{(3)}_{_{hWW}} + {\cal L}^{(4)}_{_{hhWW}}.
 \label{lag}
\end{align}
The most general effective vertices take the form:
\begin{align}
 &
 \Gamma_{hhh}
 =
 - 6 \lambda v \bigg[ \gonehhh
 + \dfrac{\gtwohhh}{3 m^2_h}
     (p_1 \cdot p_2 + p_2 \cdot p_3 + p_3 \cdot p_1) \bigg],
\\
 &
 \Gamma_{hW^-W^+}
 = g m_W
   \bigg[\bigg\{1 + \dfrac{\gonehww}{m^2_W} p_2 \cdot p_3
                + \dfrac{\gtwohww}{m^2_W} (p^2_2 + p^2_3)
         \bigg\} \eta^{\mu_2 \mu_3}
 \notag\\
 & \qquad\,\qquad\quad\quad
        - \dfrac{\gonehww}{m^2_W} p_2^{\mu_3} p_3^{\mu_2}
        - \dfrac{\gtwohww}{m^2_W} (p_2^{\mu_2} p_2^{\mu_3} + p_3^{\mu_2} p_3^{\mu_3})
 \notag\\
 & \qquad\,\qquad\quad\quad
        - {\rm i} \dfrac{\gthww}{m^2_W} \e_{\mu_2\mu_3\mu\nu} p_2^\mu p_3^\nu
   \bigg],
% \\
\end{align}
\begin{align} &
 \Gamma_{hhW^-W^+}
 = g^2
   \bigg[\bigg\{ \dfrac{1}{2}
                + \dfrac{\gonehhww}{m^2_W} p_3 \cdot p_4
                + \dfrac{\gtwohhww}{m^2_W} (p^2_3 + p^2_4)
          \bigg\} \eta^{\mu_3 \mu_4}
 \notag\\
 & \qquad\ \,\qquad\quad\quad
        - \dfrac{\gonehhww}{m^2_W} p_3^{\mu_4} p_4^{\mu_3}
        - \dfrac{\gtwohhww}{m^2_W} (p_3^{\mu_3} p_3^{\mu_4} + p_4^{\mu_3} p_4^{\mu_4})
 \notag\\
 & \qquad\ \,\qquad\quad\quad
        - {\rm i} \dfrac{\gthhww}{m^2_W} \e_{\mu_3\mu_4\mu\nu} p_3^\mu p_4^\nu
   \bigg].
\end{align}
The momenta and indices considered above are of the same order as they appear
in the index of the respective vertex $\Gamma$. For example, in the vertex
$\Gamma_{hW^-W^+}$ the momenta of $h, W^-$ and $W^+$ are $p_1,p_2$ and $p_3$
respectively. Similarly, $\mu_2$ and $\mu_3$ are the indices of $W^-$ and $W^+$.
Using the above effective field theory (EFT) approach a study has been performed as an
example for di-Higgs production in vector boson fusion at the LHC in Ref.~\citep{Alloul:2013naa}. 
%%%%%%%%%%%%%%%%%%%%%%%%%%%%%%%%%%%%%%%%%%%%%%%%%%%%%%%%%%%%%%%%%%%%%%%%%%%%%%%
\section{Simulation Tools and Analysis}
\label{sec:analysis}
We begin our probe of the sensitivity of these couplings by building a model file
for the Lagrangian in \cref{lag} using \texttt{FeynRules}~\citep{Alloul:2013naa},
and then simulate the charged current double Higgs boson production channel
$p\,e^- \to hhj\nu_e$ (see \cref{fig:figW}), with $h$ further decaying
into a $b \bar b$ pair,\footnote{In $pp$ collider like the LHC, the main
challenge of this search is to distinguish the signal of four bottom quarks in the final state
(that hadronise into jets ($b$-jets)) from the QCD multijet backgrounds. Such challenges and
difficulties are discussed and performed in ATLAS and CMS 
studies~\cite{Aad:2015uka, Aad:2015xja, Khachatryan:2015yea}. } 
in the FCC-he set up with $\sqrt{s} \approx 3.5$~TeV.
Our analysis starts with optimising the SM di-Higgs signal events with respect
to all possible backgrounds from multi-jet events, $ZZ$+jets, $hb\bar b$
+ jets, $hZ$ + jets and $t\bar t$+jets in charged and neutral current 
deep-inelastic scattering (DIS) and
in photo-production\footnote{
  We cross checked the modelling of photo-production cross sections from
  \texttt{MadGraph5} by switching on the ``Improved Weizs\"acker-Williams
  approximation formula'' described in Ref.~\citep{Budnev:1974de} to give the
  probability of photons from the incoming electron, versus the expectation of
  the \texttt{Pythia} Monte Carlo generator.
}, taking into account appropriate $b$-tagged jets and a high performance
multipurpose $4\pi$ detector. In \cref{tab:xsec} we have given an estimation
of cross sections for signal and
backgrounds considering all possible modes with basic cuts.
We then investigate the limits on
each coupling taking BSM events as the signal. For the generation of events
we use the Monte Carlo event generator \texttt{MadGraph5}~\citep{Alwall:2011uj}
and the \texttt{CTEQ6L1}~\citep{Pumplin:2002vw} parton distribution functions.
Further fragmentation and hadronisation are done with a {\em customised}
\texttt{Pythia-PGS}\footnote{
  In \texttt{Pythia-PGS} we modified several parameters in a way to use it
  for $e^-p$ collision and to get all required numbers of events demanded in
  each simulation. The coordinate system is set as for the HERA experiments, i.e.
  the proton direction defines the forward direction. The modifications have
  been successfully validated using neutral current DIS events and switched
  off QCD ISR. For $e^-p$ collisions multiple interactions and pile-up are
  expected to be very small and are switched off in our studies.
}~\citep{Sjostrand:2006za}. The detector level simulation is performed with
reasonably chosen parameters using \texttt{Delphes}\footnote{
  For \texttt{Delphes} we used the ATLAS set-up with the modifications in the
  $|\eta|$ ranges for forward and $b$-tagged jets up to 7 and 5 respectively
  with $70$\% tagging efficiency of $b$-jets as mentioned in the text. The
  resolution parameters for energy deposits in the calorimeters are based on the
  ATLAS Run-1 performance.
}~\citep{deFavereau:2013fsa} and jets were clustered using \texttt{FastJet
}~\citep{Cacciari:2011ma} with the anti-$k_T$ algorithm~\citep{Cacciari:2008gp}
using the distance parameter, $R = 0.​4​$. The factorisation and
renormalisation scales for the signal simulation are fixed to the Higgs boson
mass, $m_h = 125$~GeV. The background simulations are done with the default
\texttt{MadGraph5} dynamic scales. The $e^-$ polarisation is assumed to be
$-80$\%.
\begin{table}[!htbp]
\centering
\resizebox{\linewidth}{!}{
{\tabulinesep=5pt
\begin{tabu}{|lccc|}\hline\hline
Process                       & {\scshape cc} (fb)    & {\scshape nc} (fb)    & {\scshape photo} (fb)  \\ \hline\hline
Signal:                       & $2.40 \times 10^{-1}$ & $3.95 \times 10^{-2}$ & $3.30 \times 10^{-6}$ \\ \hline\hline
$b\bar b b\bar b j$:          & $8.20 \times 10^{-1}$ & $3.60 \times 10^{+3}$ & $2.85 \times 10^{+3}$ \\ \hline
$b\bar bjjj$:                 & $6.50 \times 10^{+3}$ & $2.50 \times 10^{+4}$ & $1.94 \times 10^{+6}$ \\ \hline
$ZZj$ ($Z\to b\bar b$):       & $7.40 \times 10^{-1}$ & $1.65 \times 10^{-2}$ & $1.73 \times 10^{-2}$ \\ \hline
$t\bar tj$ (hadronic):        & $3.30 \times 10^{-1}$ & $1.40 \times 10^{+2}$ & $3.27 \times 10^{+2}$ \\ \hline
$t\bar tj$ (semi-leptonic):   & $1.22 \times 10^{-1}$ & $4.90 \times 10^{+1}$ & $1.05 \times 10^{+2}$ \\ \hline
$hb\bar bj$ $(h \to b\bar b)$:& $5.20 \times 10^{-1}$ & $1.40 \times 10^{ 0}$ & $2.20 \times 10^{-2}$ \\ \hline
$hZj$ $(Z,h\to b\bar b)$:     & $6.80 \times 10^{-1}$ & $9.83 \times 10^{-3}$ & $6.70 \times 10^{-3}$ \\ \hline\hline
\end{tabu}}
}
\caption{\small Cross sections of signal and backgrounds in charged current
                ({\scshape cc}), neutral current ({\scshape nc}) and photo-production ({\scshape photo})
                modes for $E_e = 60$ GeV and $E_p = 50$ TeV, where $j$ is light quarks and gluons. For
                this estimation we use basic cuts $|\eta| \le 10$ for light-jets, leptons and $b$-tagged
                jets, $p_T \ge 10$ GeV, $\Delta R_{\rm min} = 0.4$ for all particles. And electron
                polarisation is taken to be $-0.8$.}
\label{tab:xsec}
\end{table}
\begin{figure*}[ht]
\noindent\begin{minipage}{\linewidth}

\centering
\resizebox{\linewidth}{!}{
{\tabulinesep=5pt
\begin{tabu}{|c|c|c|c|c|c|c|c|c|c|}\hline\hline
Cuts / Samples    & Signal  & $4b$+jets & $2b$+jets
                                                   & Top     & $ZZ$    &  $b\bar{b} H$
                                                                                & $ZH$    & Total Bkg & Significance
                                                                                                              \\ \hline\hline
Initial           & $2.00\times 10^3$
			&$3.21\times 10^7$
				&$2.32\times 10^9$
					&$7.42\times 10^6$
						&$7.70\times 10^3$
							&$1.94\times 10^4$
								&$6.97\times 10^3$
									&$2.36\times 10^9$
										&0.04\\ \hline
At least $4b+1j$  &$3.11\times 10^2$
				&$7.08\times 10^4$
					&$2.56\times 10^4$
						&$9.87\times 10^3$
							&$7.00\times 10^2$
								&$6.32\times 10^2$
									&$7.23\times 10^2$
										&$1.08\times 10^5$
											&0.94\\ \hline
Lepton rejection $p_T^\ell > 10$~GeV
                  &$3.11\times 10^2$
                  	&$5.95\times 10^4$
                  		&$9.94\times 10^3$
					&$6.44\times 10^3$
						&$6.92\times 10^2$
							&$2.26\times 10^2$
								&$7.16\times 10^2$
									&$7.75\times 10^4$
										&1.12\\ \hline
Forward jet  $\eta_J > 4.0$
                  & 233	& 13007.30	& 2151.15		& 307.67	& 381.04	& 46.82	& 503.22	& 16397.19	& 1.82	\\ \hline
$\slashed E_{T} > 40$~GeV 
                  &155	& 963.20	& 129.38	& 85.81	& 342.18	& 19.11	& 388.25	& 1927.93		& 3.48	\\ \hline
$\Delta\phi_{\slashed E_T j} > 0.4$ 
                  & 133	& 439.79	& 61.80	& 63.99	& 287.10	& 14.53	& 337.14	& 1204.35		& 3.76	\\\hline
$m_{bb}^1 \in [95,125]$, $m_{bb}^2 \in [90,125]$
                  & 54.5	& 28.69	& 5.89	& 6.68	& 5.14	& 1.42	& 17.41	& 65.23	& 6.04 \\ \hline
$m_{4b} >290$~GeV 
		& 49.2	& 10.98	& 1.74	& 2.90	& 1.39	& 1.21	& 11.01	& 29.23	&7.51 \\ \hline\hline
\end{tabu}}
}
\captionof{table}{\small A summary table of event selections to optimise the signal with respect to the backgrounds
                in terms of the weights at 10~\abinv. In the first column the selection criteria are given
                as described in the text. The second column contains the weights of the signal process
                $p\,e^- \to hhj\nu_e$, where both the Higgs bosons decay to $b\bar b$ pair. In the
                next columns the sum of weights of all individual prominent backgrounds in charged current, neutral
                current and photo-production are given with each selection, whereas in
                the penultimate column all backgrounds' weights are added. The significance is
                calculated at each stage of the optimised selection criteria using the formula ${\cal S}
                = \sqrt{2 [  ( S + B ) \log ( 1 + S/B ) - S ]}$, where $S$ and $B$
                are the expected signal and background yields at a luminosity of 10~\abinv respectively.
                This optimisation has been performed for $E_e = 60$~GeV and $E_p = 50$~TeV.}
\label{tab:cut_flow}

\bigskip

  \includegraphics[trim=0 0 0 50,clip,width=0.33\textwidth,height=0.3\textwidth]
                              {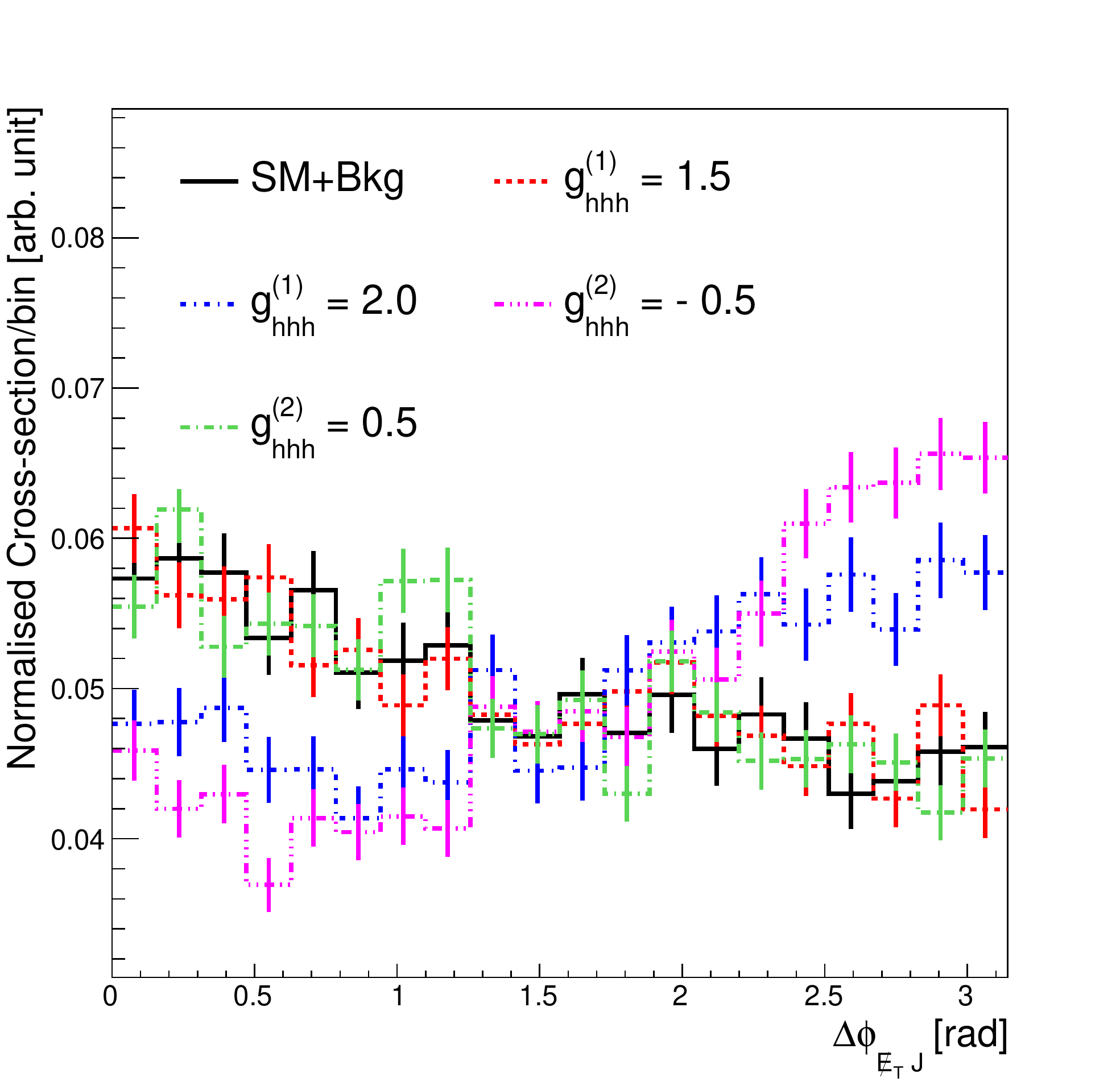}
  \includegraphics[trim=0 0 0 50,clip,width=0.33\textwidth,height=0.3\textwidth]
                              {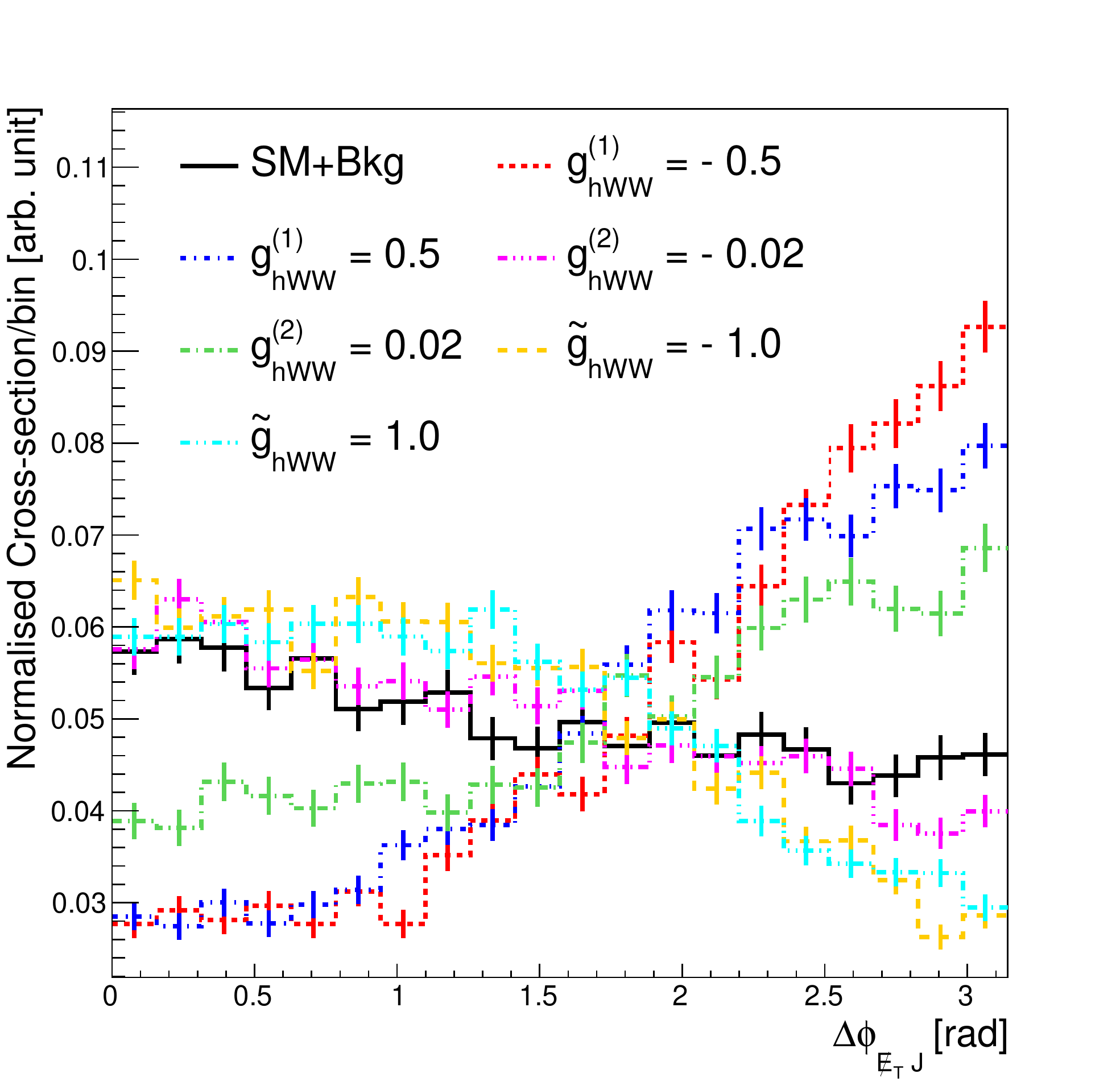}
  \includegraphics[trim=0 0 0 50,clip,width=0.33\textwidth,height=0.3\textwidth] 
                              {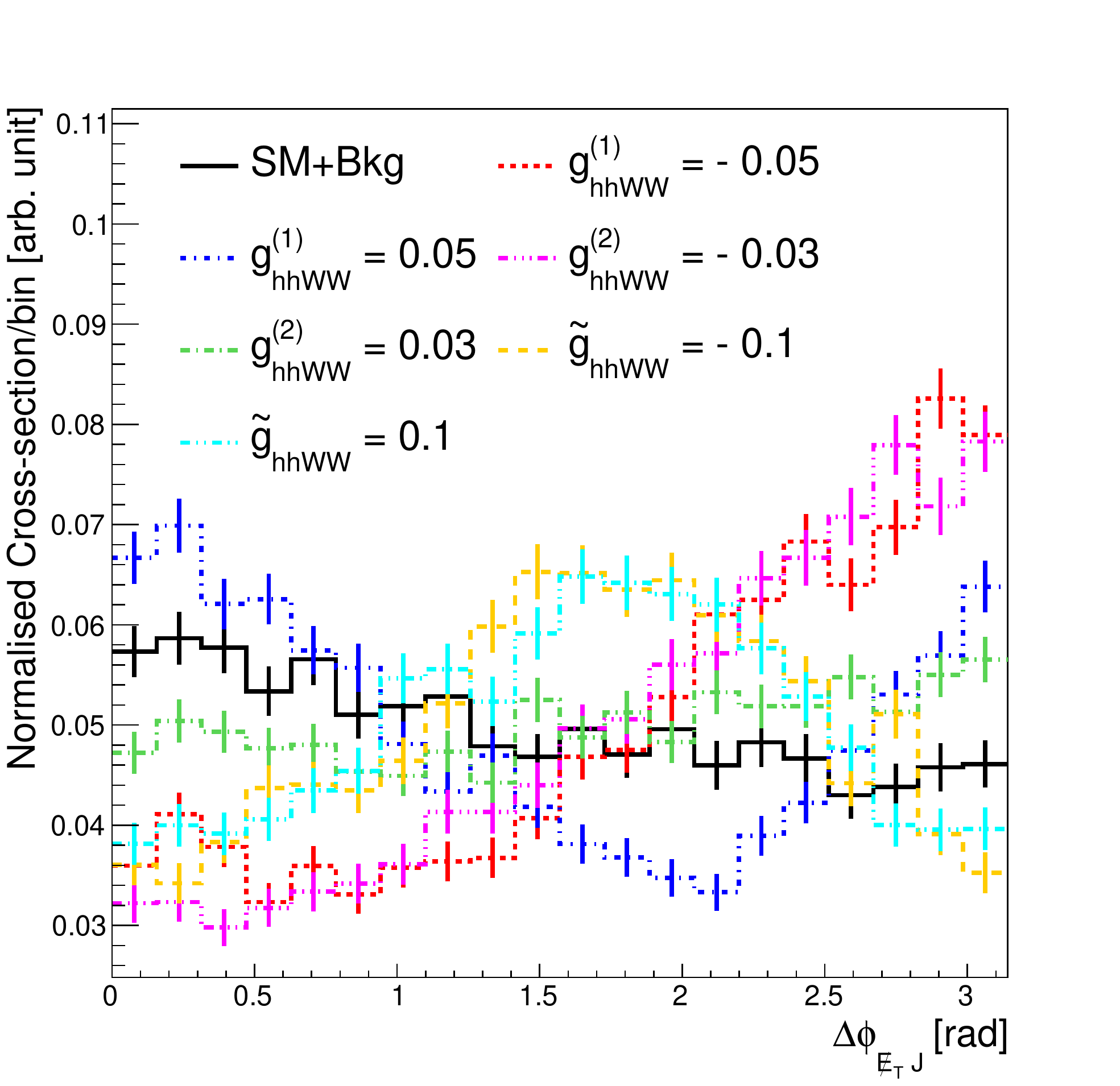}
  \captionof{figure}{\small Azimuthal angle distributions, at \texttt{Delphes} detector-level,
                  between missing transverse energy, $\slashed E_T$, and the forward
                  jet, J, in the SM (including backgrounds) and with the anomalous
                  $hhh$, $hWW$ and $hhWW$ couplings. The error bars are statistical.}
\label{fig:dist}

\end{minipage}
\end{figure*}
\subsection{Cut-based optimisation}
\label{sec:cuts}
We base our simulation on the following kinematic selections in order to optimise
the significance of the SM signal over all the backgrounds:
\begin{inparaenum}[(1)]
 \item At least four $b$-tagged jets and one additional light jet are selected
       in an event with transverse momenta, $p_T$, greater than $20$~GeV.
 \item For $non$-$b$-tagged jets, the absolute value of the rapidity, $|\eta|$,
       is taken to be less than $7$, whereas for $b$-tagged jets it is less
       than $5$.
 \item The four $b$-tagged jets must be well separated and the distance between
       any two jets, defined as $\Delta R = \sqrt{(\Delta \phi)^2+(\Delta\eta)^2}$,
       $\phi$ being the azimuthal angle, is taken to be greater than $0.7$.
 \item Charged leptons with $p_T > 10$~GeV are rejected.
 \item For the largest $p_T$ forward jet J (the $non$-$b$-tagged jet after
       selecting at least four $b$-jets) $\eta_{J} > 4.0$ is required.
 \item The missing transverse energy, $\slashed E_{T}$, is taken to be greater
       than $40$~GeV.
 \item The azimuthal angle between $\slashed E_T$ and the $b$-tagged jets are: $\Delta
       \Phi_{\slashed E_{T},\ leading\,jet} > 0.4$ and $\Delta\Phi_{\slashed
       E_{T},\ sub-leading\,jet} > 0.4$.
 \item The four $b$-tagged jets are grouped into two pairs such that the
       distances of each pair to the true Higgs mass are minimised. The leading
       mass contains the leading $p_T$-ordered $b$-jet. The first pair is required
       to be within $95$-$125$~GeV and the second pair within $90$-$125$~GeV\footnote{
  Among the four $b$-tagged jets, choices of pairing have been performed via
  appropriate selection of mass window, keeping in mind to reconstruct the Higgs boson
  mass, $m_h$, in the signal as well as the $Z$-boson mass, $m_Z$, in the backgrounds.
  We choose the pair in which the quadratic sum ($m_1 -m_c$) and ($m_2-m_c$) is
  smallest, and in each mass $m_i$, mass $m_1$ has the largest $p_T$ $b$-jet,
  $m_c = (m_h - m_0)$~GeV, and normally $m_0\approx 20$-$40$~GeV (which is not important,
  since the false pairing will have a much higher quadratic sum).
       }.
 \item The invariant mass of all four $b$-tagged jets has to be greater than
       $290$~GeV.
\end{inparaenum}
 
In the selections (described above) the $b$-tagging efficiency is assumed to be $70$\%, with fake
rates from $c$-initiated jets and light jets to the $b$-jets of $10$\% and $1$\%
respectively. Corresponding weights\footnote{Here weights mean the number of expected
events at a particular luminosity. The number of events of the photo-production
of $4b$+jets are derived using the efficiencies of the Monte Carlo samples due
to the low statistics. The other backgrounds are obtained directly from the event
selection.} at a particular luminosity of 10~\abinv for a signal, and all
backgrounds with significance has been tabulated in \cref{tab:cut_flow}. 
Significance at all stages of the cuts are calculated using the Poisson formula
${\cal S} = \sqrt{2 [  ( S + B ) \log ( 1 + S/B ) - S ]}$, where $S$ and $B$
are the expected signal and background yields at a particular luminosity respectively.

From \cref{tab:cut_flow} we recognise that selection on the forward jet in the
FCC-he type machine plays a very significant role in distinguishing the signal
with respect to background. By selecting events with $\eta_J > 4.0$, there is
loss of 25\% of signal events, while the total background loss is around 80\%.
The next significant cut on missing energy ($\slashed{E}_T > 40$~GeV) is also
very significant as due to this cut there is loss of 88\% of events in total
background, however, for the signal there is a loss of only 34\% of events after
forward jet selection. Furthermore the mass window cut for the invariant mass
of two $b$-tagged jets, after the $\Delta\phi_{\slashed E_Tj} > 0.4$ selection,
significantly reduces the total background events to 5\%, only while the signal
events remains at 40\%. Efficient requirements on the invariant mass window of
four $b$-jets are efficient, such that to reduce backgrounds by 44\% leads to a
signal of 90\% with respect to the previous two $b$-tagged jet mass window selection.
And hence there is a 20\% enhancement in the significance obtained in comparison
to the two different mass window selection criteria, and overall with respect to
initial events this cut-based optimisation is enhanced from a 0.04 to 7.51
significance. Here it is also important to mention that photo-production of
the 4$b$ final state is one of the main background with similar topological
final states from $Z h$, where $Z, h \to b\bar b$, and is equally important.
Hence choice of efficient selection criteria is too important to reduce these
backgrounds.

\subsection{Kinematic distributions and observable}
\label{kinobs}
For our analysis we take {\em ad hoc} values of positive and negative couplings
in such a manner that the production cross section does not deviate much from
the SM value, and in particular modifications in the shapes of the normalised
azimuthal angle distribution between the missing transverse energy and the
leading (forward) jet are studied, in addition to other kinematic distributions.

Taking into account all the above criteria we study BSM modifications in various
differential distributions at the \texttt{Delphes} detector-level. This leads to
the following observations:
\begin{inparaenum}[(1)]
 \item $p_T$ has the usual tail behaviour, i.e. the number of events are more
       populated in the higher $p_T$ region with respect to the SM for the
       chosen values of the anomalous couplings.
 \item In cases of the $\eta$ distributions:
       \begin{inparaenum}[(a)]
        \item For the forward jet, particularly for the couplings of $hWW$ and
              $hhWW$ vertices, the mean $\eta$ is more central in the detector.
              The behaviour is similar if we increase the c.m.s. energy of the
              collider by increasing $E_e$ to higher ($>60$~GeV) values. For
              $hhh$ couplings the $\eta$ distribution remains the same as for
              the SM.
        \item In case of $b$-tagged jets, for all values of anomalous couplings,
              the distribution is populated around the value of $\eta$ of the SM
              distribution.
       \end{inparaenum}
 \item For the specific observable of the azimuthal angle difference
       between missing transverse energy and the forward jet ($\Delta\phi_{\slashed E_T J}$)
       the shapes are clearly distinguishable from the SM.
\end{inparaenum}
This behaviour is shown in \cref{fig:dist}, where the values of the couplings
are {\em ad hoc}. However, these values are taken only for the purpose of
illustration, and in the limit of the couplings going to their SM values the
shapes will coincide with the SM distributions. The specific characteristics
of the curves also depend on the details of the selection requirements, but
the qualitative differences could be seen at every selection step. The shape
of the curves is due to the fact that all new physics couplings 
have a momentum dependent structure (apart from \gonehhh) and positive
or negative interference with SM events. 
We note that $\Delta\phi_{\slashed E_T J}$ 
is a novel observable and commands more focused and deeper analyses. In this 
regard one should follow the analysis (as performed in Ref.~\citep{Dutta:2013mva}) based
on an asymmetry with two equal bins in $\Delta\phi_{\slashed E_T J} \lessgtr \pi/2$,
defined as
\begin{gather}
 {\cal A}_{\Delta\phi_{\slashed E_T J}}
 =
 \dfrac{|A_{\Delta \phi > \pi/2}| - |A_{\Delta \phi < \pi/2}|}
       {|A_{\Delta \phi > \pi/2}| + |A_{\Delta \phi <\pi/2}|},
 \label{asym_def}
\end{gather}
\begin{table}[!htbp]
\centering
{\tabulinesep=3pt
\begin{tabu}{|l|l|c|c|}\hline\hline
 \multicolumn{2}{|c|}{Samples}                  & ${\cal A}_{\Delta\phi_{\slashed E_T J}}$ & $\sigma {(\rm fb)}$
                                                                       \\ \hline\hline
 \multicolumn{2}{|c|}{SM+Bkg}                   & $0.277 \pm 0.088$   & {} \\ \hline
\multirow{2}*{$\gonehhh$}   & = \ \ 1.5   & $0.279 \pm 0.052$    & $ 0.18$ \\
                                 & = \ \ 2.0   & $0.350 \pm 0.053$    & $ 0.21$ \\ \hline
\multirow{2}*{$\gtwohhh$}   & =   - 0.5   & $0.381 \pm 0.050$   & $ 0.19$  \\
                                 & = \ \ 0.5   & $0.274 \pm 0.024$   & $0.74 $  \\ \hline
\multirow{2}*{$\gonehww$}   & =   - 0.5   & $0.506 \pm 0.022$   & $ 0.88$  \\
                                 & = \ \ 0.5   & $0.493 \pm 0.020$   & $0.94 $  \\ \hline
\multirow{2}*{$\gtwohww$}   & =   - 0.02  & $0.257 \pm 0.025$  & $ 0.67$   \\
                                 & = \ \ 0.02  & $0.399 \pm 0.040$    & $ 0.33$ \\ \hline
\multirow{2}*{$\gthww$}     & =   - 1.0   & $0.219 \pm 0.016$    & $ 1.53$ \\
                                 & = \ \ 1.0   & $0.228 \pm 0.016$  & $1.53 $   \\ \hline
\multirow{2}*{$\gonehhww$}  & =   - 0.05  & $0.450 \pm 0.033$    & $ 0.52$ \\
                                 & = \ \ 0.05  & $0.254 \pm 0.029$   & $ 0.68$  \\ \hline
\multirow{2}*{$\gtwohhww$}  & =   - 0.03  & $0.462 \pm 0.022$  & $1.22 $   \\
                                 & = \ \ 0.03  & $0.333 \pm 0.018$   & $ 1.46$  \\ \hline
\multirow{2}*{$\gthhww$}    & =   - 0.1   & $0.351 \pm 0.020$    & $ 1.60$ \\
                                 & = \ \ 0.1   & $0.345 \pm 0.020$    & $1.61 $ \\ \hline\hline
\end{tabu}}
\caption{\small Estimation of the asymmetry, defined in \cref{asym_def}, and statistical
                error associated with the kinematic distributions in \cref{fig:dist}
                at an integrated luminosity of 10~\abinv. The cross section ($\sigma$) for the 
                corresponding coupling choice is given in the last column with same 
                parameters as in \cref{tab:xsec}.}
\label{tab:asym_dphi}
\end{table}
in which $A$ is the yields obtained for the given \abinv data after all of the
selections, including both the signal and backgrounds. \cref{tab:asym_dphi} shows
the estimation of the asymmetry for a set of representative values of the couplings,
shown in \cref{fig:dist}, along with the associated statistical uncertainty. Though
the new physics couplings are representative, we can infer the sensitivities of
these couplings with respect to the SM+Bkg estimation of asymmetry from \cref{tab:asym_dphi},
where $\gonehww$ seems to have large fluctuations for both positive and negative
choices of its values. Similarly the sensitivities of $\gtwohhww$ and $\gthhww$
can also be noted, however $\gonehhww$ is more sensible for the negative choice
of its value. The study of the sensitivity of the non-standard couplings through
this asymmetry observable considering a kinematic distribution is basically corresponding
to two bins by dividing the whole distribution in two halves with large bin-width.
Moreover, this kind of study can be further appended with finer bin widths and a
more efficient $\chi^2$ analysis (for example in Ref.~\citep{Dutta:2013mva}).
However, these detailed analyses are beyond the scope of this article.

\subsection{Exclusion limits through fiducial cross section as a function of luminosity}
\label{exc}

\begin{figure}[!htbp]
  \includegraphics[trim=0 40 0 20,clip,width=0.48\textwidth,height=0.48\textwidth]{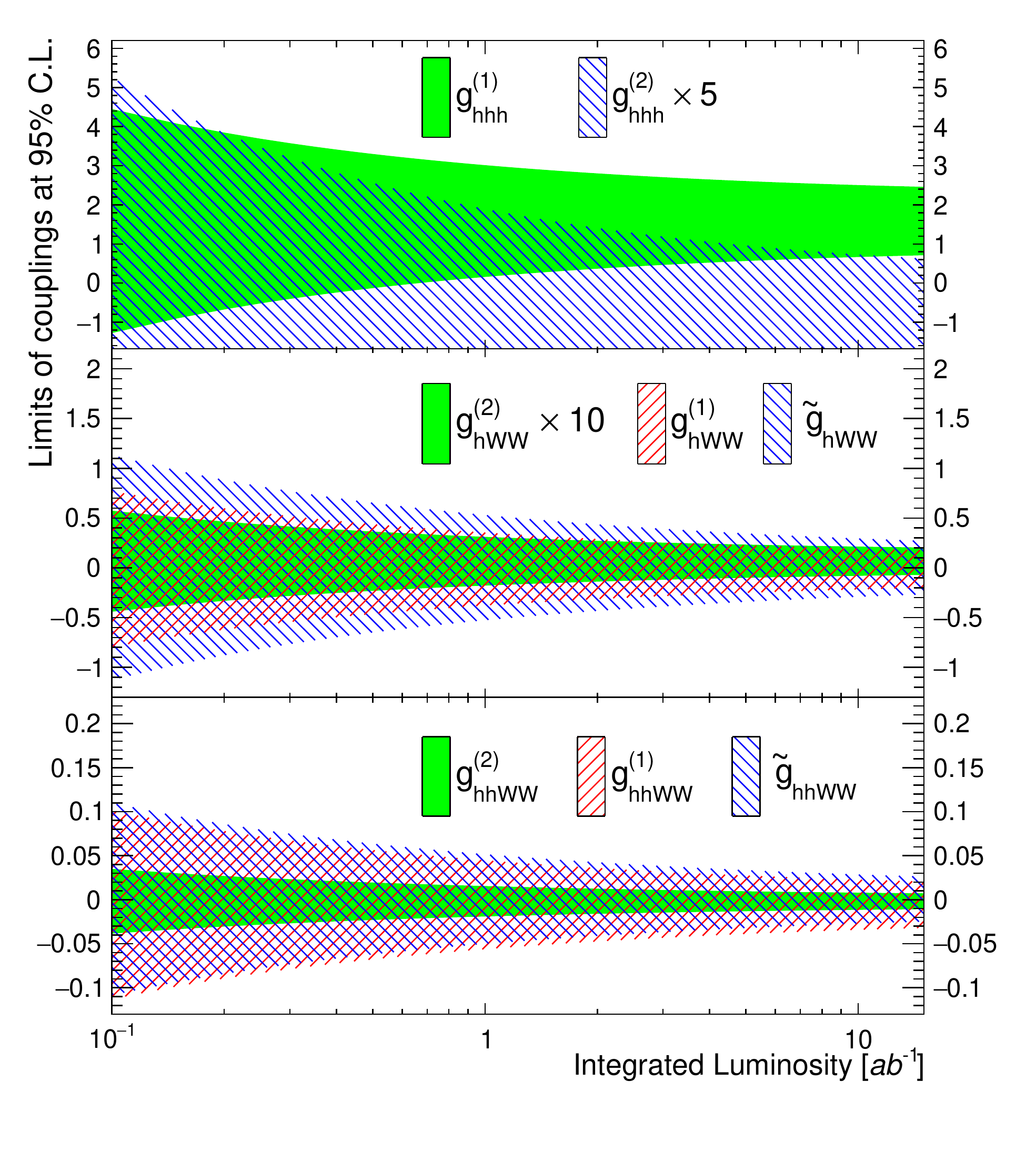}
  \caption{\small The exclusion limits on the anomalous $hhh$ (top panel), $hWW$
                  (middle panel) and $hhWW$ (lower panel) couplings at $95$\% C.L.
                  as a function of integrated luminosity (shaded areas).
                  Note that the allowed values
                  of $\gtwohhh$ and $\gtwohww$ are multiplied by $5$ and
                  $10$ respectively to highlight their exclusion region,
                  since the values are of the order $10^{-1}$.}
\label{fig:lumi}
\end{figure}

Furthermore, we probe the exclusion limits on these couplings as a function of
the integrated luminosity, with the log-likelihood method described in
Ref.~\citep{Cowan:2010js}, using directly the fiducial inclusive cross section as
an observable. In \cref{fig:lumi} we present exclusion plots at $95$\% C.L.
for anomalous $hhh$, $hWW$ and $hhWW$ couplings, where the shaded areas are
the allowed region. The exclusion limits are based on the SM `di-Higgs signal
+ backgrounds' hypotheses considering BSM contributions as the signal at the
given luminosity. Each limit is given by scanning one coupling and fixing the
other couplings to their SM value, where a $5$\% systematic uncertainty is taken
into account on the signal and background yields respectively.
From \cref{fig:lumi} our observations are as follows:
\begin{inparaenum}[(1)]
 \item If the integrated luminosity exceeds $0.5\ \abinv$ $\gonehhh$ is
       restricted to be positive. $\gonehhh$ is allowed to be within $0.7$-$2.5$
       when the integrated luminosity reaches $15~\abinv$ as for
       values of $1 < \gonehhh \leq 2.1$ the cross section is smaller than the
       SM di-Higgs production.
 \item The $\gtwohhh$ value is restricted to around $10^{-1}$. We only exclude
       the positive part of this coupling because its negative part has cancellations
       with the SM di-Higgs cross section.
 \item The sensitivity for $hWW$ couplings, namely $\gonehww$ and $\gthww$, can
       be better probed at much lower energies and luminosity at the LHeC using the
       single Higgs boson production as shown in Ref.~\citep{Biswal:2012mp}.
       However, we have shown the sensitivity of $\gtwohww$, which is not
       considered in Ref.~\citep{Biswal:2012mp}, and it is of the order $10^{-2}$
       in the allowed region.
 \item One important aspect of di-Higgs production in this type of collider is
       that one can measure the sensitivity of the $hhWW$ couplings also. In our
       analysis, since the CP-even (odd) coupling $\gonehhww$ ($\gthhww$) has
       similar Lorentz structures, with the sensitivity of the exclusion plot having
       almost the same order of magnitude. However, the structure of $\gtwohhww$
       allows a comparatively narrower region of values.
\end{inparaenum}
The couplings belonging to both the $hWW$ and $hhWW$ vertices are strongly
constrained because of their high production cross section at very low values
of the couplings. By increasing the luminosity from $0.1$-$1~\abinv$ the
constraint on the couplings increases and its limits are reduced by a factor
two. A further increase of the luminosity will not change the results.
All limits are derived by varying only one coupling at a time, as mentioned
earlier.
The exclusion limits on the couplings in this analysis are based on the constraints 
from an excess above the SM expectation while potential deficits from interference 
contributions are not sensitive yet to be used for limit settings.
    
\subsection{Prospects at higher $E_e$ and sensitivity of the Higgs self coupling}
\label{sensh}

Finally we discuss what happens once the electron energy $E_e$ is increased
to higher values, where we focus our analysis on a determination of the SM Higgs
self coupling, assuming no further BSM contributions. Without going into detail
we can note that with increasing $E_e$ (from $60$~GeV to $120$~GeV) the SM
signal and dominant background production cross sections are enhanced by a factor
of $2.2$ and $2.1$ respectively. As a result, the cut efficiency for the selection
of four $b$-tagged jets and one forward jet is improved, but for the other cuts
described previously (invariant mass, $\slashed E_T$, $\eta_{J}$ and $\Delta
\phi_{\slashed E_T j}$) it remains very similar. This leads to an enhancement
of the selected signal and dominant background events by a factor $2.5$ and
$2.6$ respectively. Hence we would obtain the same statistical precision
with only $40$\% of the luminosity of an $E_e = 60$~GeV beam when increasing the
electron energy to $120$~GeV. At an ultimate integrated luminosity of $10~\abinv$,
increasing $E_e$ from $60$ to $120$~GeV would increase the significance of the
observed SM di-Higgs events from $7.2$ to $10.6$, obtained from a
log-likelihood fit. This includes a $5$\% signal and background systematics
mentioned earlier. For the SM Higgs boson self coupling, where the scaling
factor is expected to be $\gonehhh = 1$, we perform an intelligent
signal injection test, which gives locally measured uncertainties for $\gonehhh$.
From this test the $1\sigma$ error band around the expected SM strength of this
coupling is $\gonehhh = 1.00^{+0.24(0.14)}_{-0.17(0.12)}$ for $E_e = 60 (120)$~GeV.

\section{Validity of EFT}
\label{eftval}
In the EFT-based approach for our analyses, the usual SM Lagrangian is supplemented
by higher-dimensional operators that parametrise the possible effects of new %non-observed
states assumed to appear at energies larger than the effective scales identified with
$m_W$ (or equivalently $v$) by restricting the operators of dimension less than or equal
to six. We have estimated the sensitivity of the involved coupling coefficients appearing
in the effective Lagrangians in \cref{lagh,lag3,lag4} with the EW scale
for the derivative terms. A detailed discussion with general couplings and mass scales
with a higher dimension EFT Lagrangian can be found in Ref.~\cite{Giudice:2007fh}.
For the processes at high energy, it is well known that an EFT approach provides an accurate
description of the underlying new physics as long as the energies are below the new physics
scale, $\Lambda$, and thus the limits on the couplings obtained in the above analyses
shall degrade at scales higher than the EW scales (since for the fixed values
of the couplings, the interference and pure BSM terms always give low contributions in
the cross section measurements for high values of scale choice). Also for ${\cal O}(1)$
values of anomalous couplings (apart from \gonehhh) $g^{(i)}_{(\cdots)}$, $\tilde g_{(\cdots)}$ and 
TeV-scale momenta, one reaches the regime where the operators in \cref{lagh,lag3,lag4} may
not be dominant, and operators with four and more derivatives may be equally
important. In other words, the EFT behind these Lagrangian expansions breaks down.
It would be important then to know how much the projected sensitivity depends on
events that violate this EFT bound. With an EW precision test in Ref.~\cite{Farina:2016rws},
it is shown how an EFT's reach deteriorates when only data below the cutoff scales
are employed on the mass variable in the case of Drell-Yan processes at the LHC.

\begin{figure}[!ht]
  \includegraphics[trim=0 40 0 20,clip,width=0.48\textwidth,height=0.48\textwidth]{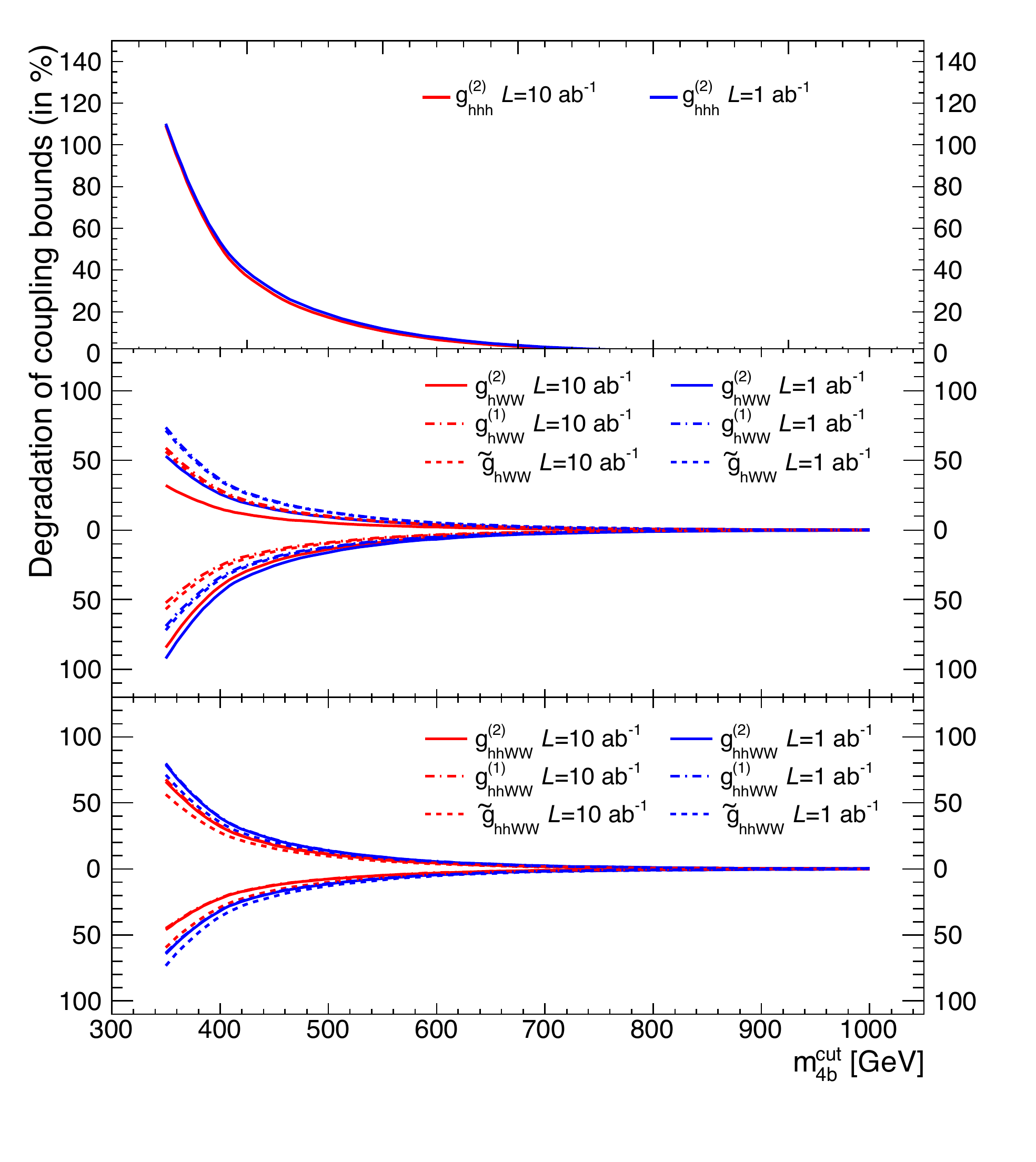}
  \caption{\small Percentage of deterioration of exclusion limits of anomalous tensorial couplings
                  (shown in \cref{fig:lumi}) with respect to the upper di-Higgs
                  invariant mass cut $m_{4b}^\prime \equiv m^{\rm cut}_{4b}$ [in GeV]
                  for fixed luminosity of 1~\abinv ({\it blue}) and 10~\abinv ({\it red}). The numbers
                  in the vertical axis above (below) 0 is the degradation in the upper
                  (lower) limits.}
\label{fig:eft_exc}
\end{figure}
   
A similar exercise can be performed in our case to estimate the deterioration of
limits on anomalous tensorial couplings $g^{(i)}_{(\cdots)}$ and $\tilde g_{(\cdots)}$ 
(the coupling coefficients which corresponds to momentum-dependent Lorentz structure) 
as a function of the
cut-off scale. In this approach we put an upper cut on the di-Higgs invariant mass
($m_{4b}^\prime$)\footnote{Note that in previous subsections we used the notation
  $m_{4b}$ for the lowest cut on di-Higgs invariant mass. Here we use $m_{4b}^\prime$
  to avoid confusion since for all analyses apart from the EFT validity we selected
  the events $m_{4b} > 290$~GeV to suppress backgrounds and increase the overall
  significance. Here, to investigate the sensitivity of BSM tensorial couplings,
  we chose the events below $m_{4b}^\prime$ cuts, keeping $m_{4b} > 290$~GeV so
  that an one to one comparison can be performed.
} such that EFT-violating events ($> m_{4b}^\prime$) are cut away, and then we
estimate by how much the projected sensitivity of $\gtwohhh$, $g_{_{hWW}}^{(1,2)}$,
$g_{_{hhWW}}^{(1,2)}$ and $\gthww$, $\gthhww$ degrades with respect to their
previous limits. In \cref{fig:eft_exc} we present the percentage of deterioration
of the exclusion limits of these anomalous effective couplings by selecting events
below $m_{4b}^\prime \in [0.35, 1]$~TeV for fixed luminosity of 1~\abinv and 10~\abinv
at 95\% C.L. It is apparent from \cref{fig:eft_exc} that the deterioration in the
limits of these anomalous couplings is large for low values of the $m_{4b}^\prime$
cut, because the effective cross section decreases (which is equivalent to the
increase of the scale $\Lambda$ of the tensorial couplings) with the decrease of
the values of $m_{4b}^\prime$. Comparing the exclusion limits obtained in \cref{fig:lumi}
we observed that at $m_{4b}^\prime = 350$~GeV the percentage of deterioration in
$\gtwohhh$ is more than 100\%, while other $hWW$ and $hhWW$ couplings deteriorate
by $~ 60 - 80$\% on both upper and lower sides at 1~\abinv and 10~\abinv. After
350~GeV a sudden decrease in degradation percentage can be noticed for
$m_{4b}^\prime = 400 - 450$~GeV for all couplings. Furthermore around
500~GeV for $\gtwohhh$, it remains 18\% while others are around 10\%. Beyond a
650~GeV cut, all the couplings converge to the original value of limits obtained
in our previous analyses, as shown in \cref{fig:lumi}.

%%%%%%%%%%%%%%%%%%%%%%%%%%%%%%%%%%%%%%%%%%%%%%%%%%%%%%%%%%%%%%%%%%%%%%%%%%%%%%%
\section{Summary and conclusions}
\label{sec:conc}
We conclude that the FCC-he, with an ERL energy of $E_e \ge 60$~GeV and a proton
energy $E_p = 50$~TeV, would provide significant di-Higgs event rates, and through this
channel one can probe accurately the Higgs boson self coupling provided that
integrated luminosities of more than $1~\abinv$ may be achieved. Along with the
Higgs self coupling one can search for any BSM signal through the measurement of
the anomalous $hhWW$ contributions. One interesting feature of this type of
machine is recognised by identifying forward jets in the signal events where an
appropriate selection, as shown for our study, reduces backgrounds efficiently
around 80\% with a loss of only 25\% of signal events. Our work also shows that
$\Delta\phi_{\slashed E_Tj}$ is a very good observable for any new physics
contributions in the given channel. Estimation of an asymmetry observable in
$\Delta\phi_{\slashed E_Tj}$ for this kinematic distribution gives a preliminary
idea of sensitivities of any new non-standard couplings. The limits on each
coupling are set by measuring the observed event rate. But the asymmetry in
$\Delta\phi_{\slashed E_Tj}$ can provide more distinguishability of the new
physics, especially cancelling many potential systematics, which is helpful
to distinguish the signatures of each model. An exclusion limit with respect
to luminosity for these couplings is studied, and a signal injection test shows
the uncertainty of the Higgs self coupling around its expected SM strength.

With all these analyses we infer that the order of sensitivities of all non-standard
couplings considered for our study within most of the luminosity ranges are consistent
with the adopted methodology of asymmetry observable, and exclusion limits through
fiducial cross sections at 95\% C.L. However, at luminosities $\sim$10-15~\abinv or
higher, the method based on fiducial cross sections constrains the non-standard
couplings more tightly. 
In addition to the fiducial cross sections, the drastic change in $\Delta\phi_{\slashed E_Tj}$
shape for $\gonehhh$ around 2.0 with respect to the SM in \cref{fig:dist} (left),
suggest that using further observables like $\Delta\phi_{\slashed E_Tj}$
may significantly improve the sensitivity of BSM couplings in general.
It is to be noted that the non-standard momentum dependent
structures of the EFT breaks down at the TeV energy regime for couplings of ${\cal O}(1)$
and then additional derivative terms become relevant. 
Hence we also show the
deteriorations in the limits of anomalous tensorial couplings for different regions
of di-Higgs invariant mass (upper) cuts by an exclusion at fixed luminosity of 1~\abinv
and 10~\abinv corresponding to 95\% C.L. limits, with respect to the limits obtained 
using fiducial inclusive
cross section as an observable. This method is used as an alternative approach to
estimate the sensitivity of the scale dependent couplings in EFT and gives a probe
to understand the regions where validity of EFT breaks down.

Our studies show a unique capability and potential of FCC-he collider to probe
the precision measurement of not only the Higgs boson self coupling but also other
involved couplings with tensorial structure through di-Higgs boson production.

\section*{Acknowledgements}

We acknowledge fruitful discussions within the LHeC Higgs group, especially
with Masahiro Kuze and Masahiro Tanaka. RI acknowledges the DST-SERB grant
SRlS2/HEP-13/2012 for partial financial support.

\section*{References}

\bibliography{mybibfile}

\end{document}